\documentclass[preprint,aps,pra,superscriptaddress]{revtex4-1}

\usepackage{graphicx,siunitx,amsmath,physics,mathtools,bbold}
\usepackage[hidelinks]{hyperref}

\DeclareGraphicsExtensions{.pdf,.png,.jpg}

\begin{document}

\title{Quantum-dot based telecom-wavelength
quantum relay}

\author{J. Huwer}
\email{jan.huwer@crl.toshiba.co.uk}
\affiliation{Toshiba Research Europe Limited, Cambridge Research Laboratory, 208 Cambridge Science Park, Milton Road, Cambridge, CB4 0GZ, United Kingdom}%

\author{M. Felle}
\affiliation{Toshiba Research Europe Limited, Cambridge Research Laboratory, 208 Cambridge Science Park, Milton Road, Cambridge, CB4 0GZ, United Kingdom}%
\affiliation{Centre for Advanced Photonics and Electronics, University of Cambridge, J.J. Thomson Avenue, Cambridge, CB3 0FA, United Kingdom}

\author{R. M. Stevenson}
\affiliation{Toshiba Research Europe Limited, Cambridge Research Laboratory, 208 Cambridge Science Park, Milton Road, Cambridge, CB4 0GZ, United Kingdom}%

\author{J. Skiba-Szymanska}
\affiliation{Toshiba Research Europe Limited, Cambridge Research Laboratory, 208 Cambridge Science Park, Milton Road, Cambridge, CB4 0GZ, United Kingdom}%

\author{ M. B. Ward}
\affiliation{Toshiba Research Europe Limited, Cambridge Research Laboratory, 208 Cambridge Science Park, Milton Road, Cambridge, CB4 0GZ, United Kingdom}%

\author{I. Farrer}
\altaffiliation[Present address: ]{Department of Electronic \& Electrical Engineering, University of Sheffield, Sheffield, S1 3JD, United Kingdom}
\affiliation{Cavendish Laboratory, University of Cambridge, J.J. Thomson Avenue, Cambridge, CB3 0HE, United Kingdom}%

\author{R. V. Penty}
\affiliation{Centre for Advanced Photonics and Electronics, University of Cambridge, J.J. Thomson Avenue, Cambridge, CB3 0FA, United Kingdom}

\author{ D. A. Ritchie}
\affiliation{Cavendish Laboratory, University of Cambridge, J.J. Thomson Avenue, Cambridge, CB3 0HE, United Kingdom}%

\author{A. J. Shields}
\affiliation{Toshiba Research Europe Limited, Cambridge Research Laboratory, 208 Cambridge Science Park, Milton Road, Cambridge, CB4 0GZ, United Kingdom}%

\date{\today}

\begin{abstract}

The development of quantum relays for long haul and attack-proof quantum communication networks operating with weak coherent laser pulses requires entangled photon sources at telecommunication wavelengths with intrinsic single-photon emission for most practical implementations. Using a semiconductor quantum dot emitting entangled photon pairs in the telecom O-band, we demonstrate for the first time a quantum relay fulfilling both of these conditions. The system achieves a maximum fidelity of 94.5\,\% for implementation of a standard 4-state protocol with input states generated by a laser. We further investigate robustness against frequency detuning of the narrow-band input and perform process tomography of the teleporter, revealing operation for arbitrary pure input states, with an average gate fidelity of 83.6\,\%. The results highlight the potential of semiconductor light sources for compact and robust quantum relay technology, compatible with existing communication infrastructures.

\end{abstract}

\maketitle

\section{Introduction}

Optical quantum communication is of fundamental importance for a wide range of emerging quantum technologies. Its applications range from securing integrity of classical communication channels by means of quantum key distribution (QKD) \cite{Ekert1991PRL,Gisin2002RMP} to the link of distributed quantum information units, important for the scalability of future quantum processor frameworks \cite{Kimble2008N}.

Quantum information for quantum communication purposes is conveniently encoded in photonic qubits which consist of single photons or more generally weak coherent optical pulses \cite{Lo2005PRL} that can be transmitted over optical fiber. To minimize losses, these systems ideally operate in the standard telecom wavelength bands, enabling point-to-point communication over distances of up to a few hundred kilometers \cite{Rosenberg2007PRL,Wang2012OL}. Yet the principal laws of quantum mechanics prohibit the duplication of a single qubit, precluding amplification of quantum signals, as required for long-haul transmission in a global quantum communication network. Alternative devices have to be developed, namely quantum relays \cite{Jacobs2002PRA} and quantum repeaters \cite{Azuma2015NC,Briegel1998PRL}, to effectively reduce noise in a quantum channel and boost transmission distances. Both systems rely on the distribution of entanglement across the network, which is then employed to teleport bits of quantum information \cite{Bennett1993PRL}. Most appealing for QKD applications is that security of the quantum channel is unconditionally guaranteed \cite{Lo1999S}, even if the entanglement resource and the device itself are provided by an untrustworthy third party.

Entangled photon sources based on non-linear processes enabled pioneering work for the demonstration of photonic quantum teleportation \cite{Bouwmeester1997N}, most recently even by using  fully independent sources \cite{Sun2016NP,Valivarthi2016NP}. Nevertheless, the number statistics of emitted photon pairs from these sources follow a Poissonian distribution. This typically increases error rates, requiring operation at very low intensities or use of sophisticated security protocols \cite{Lo2005PRL}, not desirable for robust applications. In contrast, entangled photon sources based on semiconductor quantum dots (QD) \cite{Benson2000PRL,Michler2000S,Muller2014NP,Stevenson2006N} have purely sub-Poissonian statistics and can be directly electrically driven \cite{Salter2010N}, making them promising candidates for the development of practical technology required for cost-effective and efficient quantum networks with untrusted nodes.

The recent first implementation of a quantum-dot-driven quantum relay \cite{Varnava2016NPJQI} has proven the  feasibility of this approach by achieving state-of-the-art high-fidelity operation. Nevertheless, the experiment made use of a conventional QD emitting at short wavelength, suffering from high losses when being transmitted over long optical fiber and therefore not suitable for a scalable architecture. Great efforts have been put into the development of QDs emitting at telecom wavelengths \cite{Alloing2005APL,Benyoucef2013APL,Ward2005APL} culminating in independent demonstrations of entangled photon emission \cite{Ward2014NC} and interference with laser photons \cite{Felle2015APL}. Here, we report for the first time the implementation of a quantum-relay experiment in the telecommunication O-band ($\sim$1310\,nm), suitable for secure quantum communication applications and at the same time fully compatible with the existing communication infrastructure regarding networks and sources. 

\section{Experimental setup}

\begin{figure}[t]%
\includegraphics[width=8.4cm]{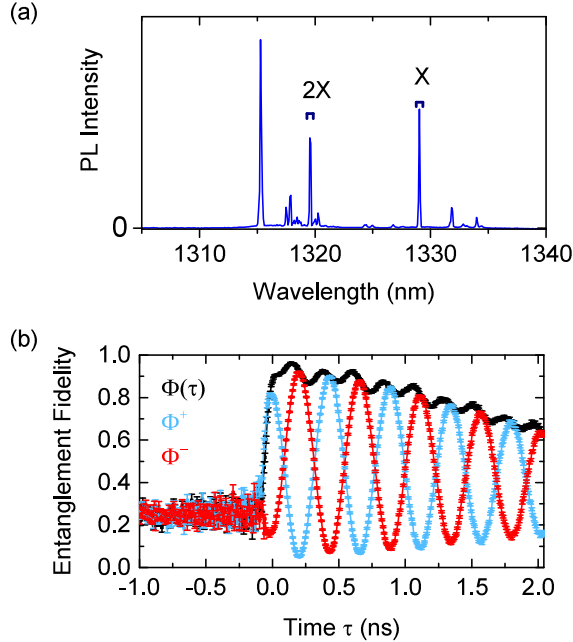}%
\caption{(a) Photo luminescence spectrum of quantum dot emission in telecom O-band with indicated recombination lines of Biexciton (2X) and Exciton (X) photons.  (b) Measured fidelity of photon pairs to the maximally entangled $\Phi^+$ (blue) and $\Phi^-$ (red) Bell states, and a time evolving state as a function of time delay $\tau$ between 2X- and X-photon emission. Error bars indicate the propagated errors from Poissonian counting statistics.}
\label{fig:dotchar}%
\end{figure}

At the core of the experiment is an entangled photon pair source based on a self-assembled InAs/GaAs QD in a \textit{p-i-n} structure, grown by molecular beam epitaxy, similar to the sample used in \cite{Felle2015APL,Ward2014NC}, but with an improved ultra-low QD density ($< \SI{0.02}{\micro\meter}^{-2}$).  The device is operated at 10\,K and optically excited with continuous-wave 1064\,nm laser light at a bias of 0\,V. The excitation power is set for equal intensity of the Biexciton (2X) and Exciton (X) spectral lines, measured with an InGaAs array in a grating spectrometer. The carefully chosen operation conditions result in a coherence time of $95\pm8$\,ps for 2X photons, measured with a Mach Zehnder interferometer. This is sufficient to guarantee indistinguishability of the photons within the timing resolution of our photon detection setup, as required for the Bell-state measurement in the teleporter. Emission from the QD is coupled into standard single-mode fiber using a confocal microscope setup (NA\,=\,0.68) and entangled photons from the 2X and X state close to the center of the telecom ‘O’-band (Fig. \ref{fig:dotchar}(a)) are spectrally isolated with a transmission grating.

Photon entanglement generated from the optical cascade of the doubly excited 2X state via emission from intermediate X levels is described by the maximally entangled state
\begin{equation}
	|\Phi(\tau)\rangle = \frac{1}{\sqrt{2}}\left( |H_{2X}H_{X}\rangle + \exp\left( \frac{iS\tau}{\hbar} \right) |V_{2X}V_{X}\rangle \right).
\label{eq:entangledstate}
\end{equation}
\textit{H} and \textit{V} denote horizontal and vertical photon polarization. \textit{S} accounts for the so-called exciton-spin splitting which is present in most QDs and related to asymmetries in their shape, resulting in a time-evolving phase of the emitted photon pair \cite{Stevenson2008PRL}. Figure \ref{fig:dotchar}(b) shows a measurement of the entanglement fidelity in our experiment, using the procedure described in \cite{Michler2009,Ward2014NC}, with the characteristic oscillatory behavior. From the measured oscillation, we derive a splitting \textit{S} of $9.05\pm\SI{0.01}{\micro\electronvolt}$. Owing to the improved QD density, we achieve a maximum fidelity of $92.0 \pm 0.2$\,\% to the static Bell $\Phi^-$ state and $96.3 \pm 0.3$\,\% to the exact time-evolving state, to our knowledge the highest value ever reported from a QD emitter.

\begin{figure}[hb]%
\includegraphics[]{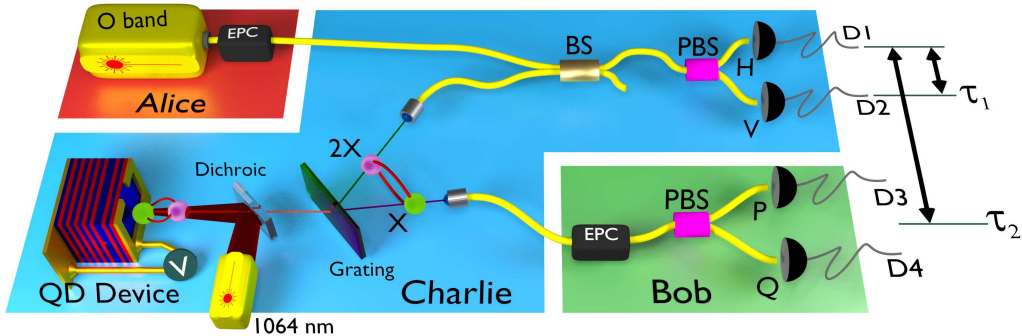}%
\caption{Quantum relay setup including QD entangled photon pair source. Alice encodes input states by use of a telecom laser and electronic polarization controller (EPC). Charlie performs a Bell-state measurement of the input and 2X photons with an unbalanced beam splitter (BS), polarizing beam splitter (PBS) and single-photon detectors D1 and D2. Bob uses an EPC, PBS and detectors D3 and D4 to analyze the output state of the X photons in different polarization bases (\textit{P}-\textit{Q}).}
\label{fig:setup}%
\end{figure}

Figure \ref{fig:setup} gives an overview of the main components of the quantum relay setup. Alice uses a commercial tuneable diode laser (linewidth 400\,kHz) with an attenuator and electronic polarization controller (EPC) to send a stream of weak coherent input qubits in different polarization bases. Spectral overlap with the 2X emission was controlled with a spectrometer and a fitting routine to overcome the limited instrument resolution, achieving an overall precision of $\pm 480$\,MHz.

At the relay station, the input mode and the 2X photon from the QD emitter are overlapped in an unbalanced 0.18\,dB coupler. Subsequent coincidence detection of photons at opposite outputs of a polarizing beam splitter (PBS) with detectors D1 and D2, corresponding to \textit{H} and \textit{V} polarization, projects the two interfering modes onto the Bell $\Psi^+$ state, heralding the teleportation of the input qubit onto the polarization state of the X photon (apart from a unitary transformation) \cite{Bennett1993PRL}.

The output is sent to the receiver (Bob) where it is analyzed in different polarization bases (\textit{PQ}), using an EPC, PBS and detectors D3 and D4. Photon detection is performed with superconducting nanowire single photon detectors \cite{SingleQuantum} and time-correlated single photon counting electronics with a measured cross-channel timing jitter of 70\,ps. Count rates only from the photon pair source during the experiments are typically $> 200$\,kcps on each detector channel for X and $> 150$\,kcps for 2X, the latter being due to higher losses in the Bell-state measurement setup at Charlie.

For all measurements, the laser intensity is adjusted to achieve 90\,\% of the detected 2X photon rate. This optimal ratio is predicted from simulations based on the semi-empirical model described in detail in \cite{Varnava2016NPJQI}, by taking all independently determined experimental parameters into account.

\section{Results}

We started characterization of the quantum relay for implementation of a standard BB84 protocol \cite{Bennett1984IEEE} with Alice sending one of the four linear polarization states \textit{H}, \textit{V}, $D=(H+V)/\sqrt{2}$, and $A=(H-V)/\sqrt{2}$. Coincidence detections between detector D1 and D2 at Charlie herald the successful operation of the relay and have to be correlated with detections of the target photons at the receiver Bob, requiring the acquisition of three-photon coincidence maps as a function of time delays $\tau_1 = t_{\textnormal{D2}} - t_{\textnormal{D1}}$ and $\tau_2 = t_{\textnormal{Bob}} - t_{\textnormal{D1}}$. For each input state, we evaluated the fidelity to the expected output.

Different physical mechanisms limit the maximum size of the temporal post selection window applicable for data evaluation. Coincidence detections along $\tau_1$ in the Bell-state measurement are sensitive to the interference visibility of the two distinct photon sources, which is influenced by the coherence time of 2X photons and timing jitter. For discrimination along $\tau_2$, between the heralding event and detection of the target, the evolution of the phase of entangled photon pairs with period $2\pi\hbar/S = 457 \pm 0.5$\,ps provides an upper limit.

\begin{figure}[ht]%
\includegraphics[width=8.4cm]{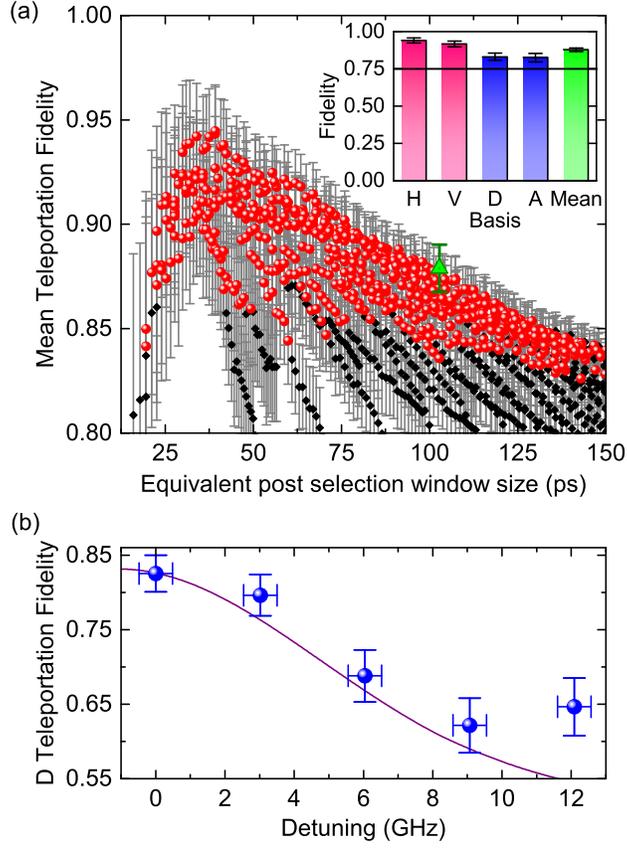}%
\caption{Measured teleportation fidelities. (a) Average teleportation fidelity for implementation of a 4-state protocol with Alice sending \textit{H}, \textit{V}, \textit{D}, and \textit{A}. Each point corresponds to a different size of the temporal post-selection window $\Delta\tau_1 \times \Delta\tau_2$ (in steps of 8\,ps), data is plotted versus the square root of the window area. Red circles denote points with all individual teleportation fidelities above 75\,\%. The inset bar plot shows individual fidelities for $\Delta\tau_1 \times \Delta\tau_2 = \SI{88}{\pico\second} \times \SI{120}{\pico\second}$, indicated by the green triangle, with values of $94.1 \pm 1.7$\,\% (\textit{H}), $91.7 \pm 1.9$\,\% (\textit{V}), $83.1 \pm 2.5$\,\% (\textit{D}), $82.5 \pm 2.8$\,\% (\textit{A}), and $87.9 \pm 1.1$\,\% (Mean). Errors are propagated from Poissonian statistics for acquired raw data. (b) Teleportation fidelity with input \textit{D} for different detuning from the center wavelength of the relay interface (2X photon). The solid line shows the corresponding simulation.}
\label{fig:4states}%
\end{figure}

Figure \ref{fig:4states}(a) shows the resulting mean teleportation fidelity for the protocol when varying the size of the post-selection window along $\tau_1$ and $\tau_2$.  The inset shows the point of highest significance, with an average teleportation fidelity of $87.9 \pm 1.1$\,\%. This exceeds the corresponding classical limit of 75\,\% by more than 11 standard deviations and is above the 80\,\% error correction threshold required for secure QKD \cite{Chau2002PRA}. When further reducing the time window at the cost of teleported qubits, we observe a maximum fidelity of $94.5 \pm 2.2$\,\% which ideally would contribute to 0.385 secure bits per detected three-fold coincidence \cite{Shor2000PRL}.

Practical implementations of a quantum relay should not only be compatible with passive telecom networking infrastructures but also with standard (non-scientific grade) telecom laser sources. These suffer from frequency drifts induced by temperature change and aging during continuous operation in non-laboratory environments. We have investigated robustness of the relay against spectral detuning of the laser-generated input. Figure \ref{fig:4states}(b) displays the measured fidelity for input \textit{D} at different detuning from the resonance of QD emission. The results show that for a frequency mismatch of 3\,GHz, which is about the linewidth of the QD emission, we can still achieve fidelities of 80\,\%, enabling secure QKD applications, and even for 6\,GHz of detuning, the fidelity exceeds the general classical limit of 2/3. The results compare well with the displayed model simulation and indicate compatibility of the system with standard telecom light sources with typical spectral stabilities in the range of 1\,GHz.

\begin{figure}[ht]%
\includegraphics[width=8.4cm]{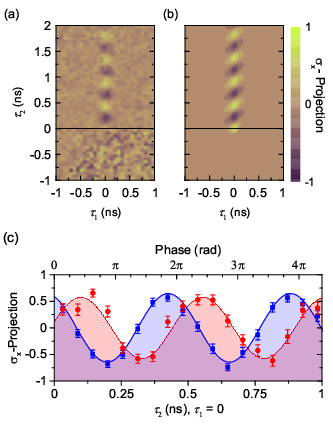}%
\caption{Temporal evolution of teleported superposition states. (a) Measured $\sigma_x$-projection for input \textit{D} with oscillation along $\tau_2 \geq 0$. (b) Corresponding simulation. (c) Cut through (a) along $\tau_2$ at $\tau_1 = 0$ for bin size $\Delta\tau_1 = 72$\,ps and $\Delta\tau_2 = 56$\,ps. Blue squares are for input \textit{D} (output \textit{D}) and red circles show the same measurement for input \textit{R} (output \textit{L}). Amplitude and phase are extracted with sinusoidal fits. There is a global fixed phase offset from $\tau_2 = 0$ due to decay from the 2X state before reaching maximum population. Error bars are propagated from Poissonian statistics of acquired raw data.}
\label{fig:statetomography}%
\end{figure}

Finally, we characterized the general performance of the implemented photonic teleporter. For arbitrary pure input states
\begin{equation}
	|\varphi_{in}\rangle = \cos\left( \frac{\theta}{2} \right) |H_L\rangle + e^{i\phi}\sin\left( \frac{\theta}{2}\right)|V_L\rangle
\label{eq:teleporterin}
\end{equation}
and under the assumption of an ideal teleportation process, we expect the output for $\tau_1 = 0$ to be of the form
\begin{equation}
	|\varphi_{out}(\tau_2)\rangle = \cos\left( \frac{\theta}{2} \right) |V_X\rangle + e^{i(\phi-S\tau_2/\hbar)}\sin\left( \frac{\theta}{2}\right)|H_X\rangle\,.
\label{eq:teleporterout}
\end{equation}
Note that in addition to a bit flip, the state is carrying the time-dependent phase acquired from the photon pair source. Figure \ref{fig:statetomography}(a) shows the resulting oscillation for input \textit{D} along $\tau_1 = 0$ in the three-photon coincidence space, measured when detecting the output at Bob in the diagonal basis ($\sigma_x$ projection). The results compare well with the displayed model calculation (Fig. \ref{fig:statetomography}(b)) and enable a direct observation of the preservation of coherence in the teleportation process.

Rather than being a limitation, the time evolving character of the output state provides an elegant way to perform quantum tomography. Full characterization of a polarization state generally requires three projective measurements along the principal axes of the Poincar\'{e} sphere \cite{James2001PRA}. The temporal evolution in equation \ref{eq:teleporterout} corresponds to a rotation of the state in the equatorial plane of the Poincar\'{e} sphere, enabling the projection onto that plane rather than a single axis, by extracting amplitude and phase of the oscillation observable in the diagonal (\textit{DA}) basis. Figure \ref{fig:statetomography}(c) compares the measured \textit{DA}-projection for inputs $R=(H+iV)/\sqrt{2}$ and \textit{D}, showing the expected phase shift of $\pi/2$ in the teleported state. Full tomography of the output at $\tau_2 = 0$ is completed by one additional projection onto the linear (\textit{HV}) basis in a separate measurement and by using the methods described in \cite{James2001PRA}. For the analysis, for each 
measurement basis a total of 1.008\,ns of data along $\tau_2$ was taken into account, corresponding to more than two full rotations of the state.

\begin{figure}[ht]%
\includegraphics[width=8.4cm]{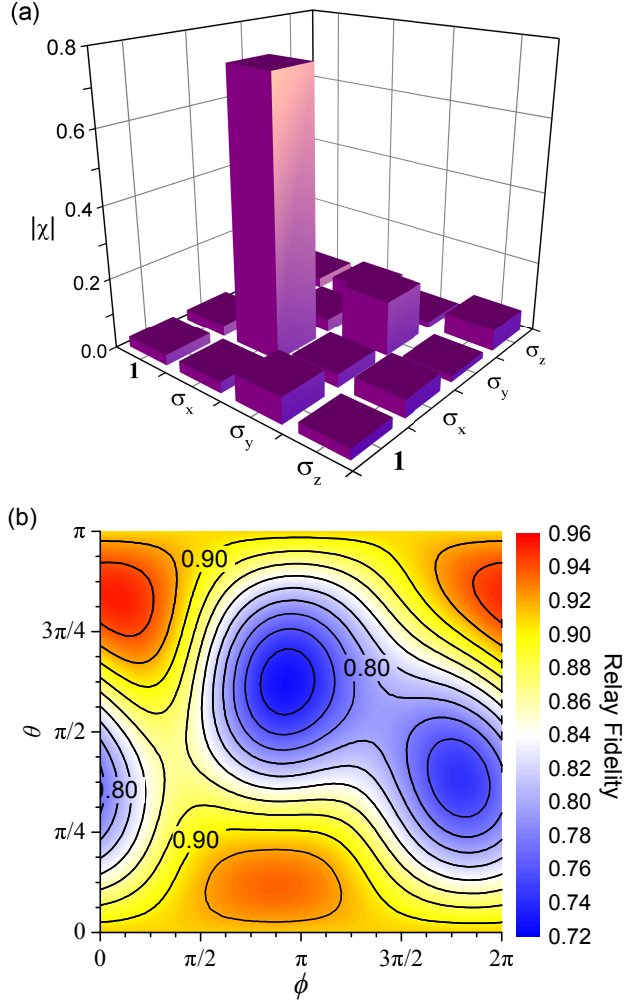}%
\caption{Quantum process tomography. (a) Modulus of reconstructed process matrix with $\sigma_x$ component of  $0.754 \pm 0.016$. (b) Calculation of relay fidelity based on the process matrix, for arbitrary pure input states parametrized by angles $\theta$ and $\phi$.}
\label{fig:processtomography}%
\end{figure}

The characterization of a quantum process can be described by the process matrix $\chi$. For single qubits, the mapping process  $\epsilon(\rho)$ for 
an input state represented by the density matrix  $\rho$ is conveniently expressed as $\epsilon(\rho)=\sum_{mn} \sigma_m \rho \sigma_n \chi_{mn}$ with $\sigma_i = \{𝟙\mathbb{1}, \sigma_x, \sigma_y, \sigma_z\}$ being the identity and Pauli operators. Using the methods described before, we did state tomography of the output for inputs \textit{H}, \textit{V}, \textit{D}, and \textit{R} enabling quantum process tomography following the procedure described in \cite{Nielsen2000Book}. Figure \ref{fig:processtomography}(a) shows the reconstructed process matrix with the dominant contribution from the bit-flip operator $\sigma_x$. We deduce a process fidelity of $75.4 \pm 1.6$\,\%, corresponding to an average teleportation gate fidelity of $83.6 \pm 1.1$\,\% \cite{Nielsen2002PLA} for arbitrary input states. It has to be emphasized that this value is to be considered as a lower bound due to the large time window along $\tau_2$ (more than 1\,ns) taken into account for the tomography.

Based on the reconstructed process matrix, we calculated the relay fidelity for arbitrary pure states as described by equation \ref{eq:teleporterin}, providing a better insight into the photonic teleportation process. Figure \ref{fig:processtomography}(b) shows the fidelity of the experimentally expected output to the corresponding ideal output state, under the assumption that the ideal teleporter is described by a pure $\sigma_x$ operation.

We observe high fidelities exceeding 90\,\% close to the polar states \textit{H} and \textit{V} ($\theta = 0, \pi$). As these correspond to the measurement basis of the Bell setup at Charlie, the gate is mainly limited by the entanglement fidelity of the photon pair source, the extinction ratio of the polarization detection optics and dark counts. The situation is rather different when approaching superposition states with $\theta = \pi/2$. Here, two-photon interference comes into play which is dominated by the Poissonian statistics of the input qubits \cite{Felle2015APL,Stevenson2013NC,Varnava2016NPJQI} and not the relay itself, lowering the teleportation fidelities. Still, minimum values are safely above the classical limit of 2/3, indicating possible quantum teleportation for arbitrary input states. Furthermore, at no point does the fidelity drop below 72.4\,\%, being the threshold required for secure implementation of 6-state protocols \cite{Bruss1998PRL,Chau2002PRA}.

\section{Conclusion}

We have demonstrated quantum teleportation of weak coherent telecom-wavelength polarization qubits making use of a semiconductor quantum-dot entangled photon-pair source. We have characterized the performance for implementation of a 4-state protocol in a quantum relay, with fidelities compatible with required security thresholds and maximum values exceeding 90\,\%. We could further show robustness of the teleportation against frequency drifts of laser-generated input states as they are expected for off-the-shelf coherent telecom light sources. The time-evolving character of the entangled photon pair source enabled us to directly observe the phase coherence of the photonic teleportation process. Finally, we took advantage of this effect for quantum state tomography of teleported states and reconstructed the process matrix of the relay, evidencing quantum teleportation for arbitrary pure input states.

The exclusive operation of the demonstrated system at telecom wavelength combined with sub-Poissonian statistics of the entangled photon pair source represents a promising platform which is in principle cascadable as required for long-haul communication in a quantum relay scheme \cite{Jacobs2002PRA}. Not only is it compatible with existing communication infrastructures, it also provides intrinsic security without the need for additional security protocols, which might prove advantageous for scalable implementation. Furthermore, the prospect of teleporting arbitrary input states shows a great potential beyond cryptography applications, for the broad spectrum of emerging photonic quantum-information technologies. The results provide solid evidence that semiconductor telecom-light sources can be considered as a serious alternative to existing well-stablished approaches for generation of photon entanglement at telecom wavelength, an important step for the development of standard quantum-relay technology for the future.

\begin{acknowledgments}
The authors acknowledge partial financial support from the Engineering and Physical Sciences Research Council, and the EPSRC Quantum Technology Hub in Quantum Communications. MF acknowledges support from the EPSRC CDT in Photonic Systems Development.

JH would like to thank M. Lucamarini for theoretical support and T. M\"uller for technical assistance.
\end{acknowledgments}

\end{document}